\documentclass{article}
\usepackage{svmult2e}
\usepackage{graphicx}
\begin{document}
\pagenumbering{arabic}

\newcommand{\req}[1]{(\ref{#1})}
\newcommand{\be}{\begin{equation}}
\newcommand{\ee}{\end{equation}}
\newcommand{\bea}{\begin{eqnarray}}
\newcommand{\eea}{\end{eqnarray}}
\newcommand{\dd}{\textrm{d}}
\newcommand{\pr}[1]{\left(#1\right)}
\newcommand{\cro}[1]{\left[#1\right]}
\newcommand{\acc}[1]{\left\{#1\right\}}
\newcommand{\e}{{\rm e}}

\def\rit{\hbox{\it I\hskip -2pt  R}}

\newcommand{\avg}[1]{\langle{#1}\rangle}
\newcommand{\ovl}[1]{\overline{#1}}
\newcommand{\BE}{\begin{eqnarray}}
\newcommand{\EE}{\end{eqnarray}}
\newcommand{\BEn}{\begin{eqnarray*}}
\newcommand{\EEn}{\end{eqnarray*}}
\newcommand{\barr}{\begin{array}}
\newcommand{\earr}{\end{array}}
\newcommand{\qe}{\`{e }}
\newcommand{\eq}{\'{e }}
\newcommand{\qa}{\`{a }}
\newcommand{\qo}{\`{o }}
\newcommand{\qi}{\`{\i }}
\newcommand{\bit}{\begin{itemize}}      
\newcommand{\eit}{\end{itemize}}
\newcommand{\bc}{\begin{center}}
\newcommand{\ec}{\end{center}}
\newcommand{\ben}{\begin{enumerate}}    
\newcommand{\een}{\end{enumerate}}
\newcommand{\nid}{\noindent}
\newcommand{\cl}{\centerline}
\newcommand{\nl}{\newline}
\newcommand{\ul}{\underline}

\newcommand{\eps}{\epsilon}
\newcommand{\de}{\partial}
\newcommand{\impl}{\Longrightarrow}
\newcommand{\To}{\longrightarrow}
\newcommand{\LRARR}{\Longleftrightarrow}

\newcommand{\Tb}{{\bf T}}
\newcommand{\Gb}{{\bf G}}
\newcommand{\Fb}{{\bf F}}
\newcommand{\xb}{{\bf x}}
\newcommand{\yb}{{\bf y}}
\newcommand{\eb}{{\bf e}}
\newcommand{\wb}{{\bf w}}
\newcommand{\ub}{{\bf u}}
\newcommand{\om}{\omega}
\newcommand{\sgn}{{\rm sgn}\,}


\title{Competition between adaptive agents:\\from learning to
collective efficiency and back}
\author{Damien Challet\\\small Theoretical Physics, Oxford University, 1 Keble Road, Oxford OX1 3NP, United Kingdom}

\maketitle

\begin{abstract}
We use the Minority Game and some of its variants to show how efficiency 
depends on learning in models of agents competing for
limited resources. Exact results from statistical physics give a clear
understanding of the phenomenology, and opens the way to the study
of reverse problems. What agents can optimize and how well is discussed in details.\\ 
\end{abstract}

Designed a simplification of Arthur's El Farol bar problem \cite{Arthur}, the
Minority Game \cite{CZ97,web} provides a natural framework for
studying how selfish adaptive agents can cope with competition. The major
 contribution of the Minority Game is not only to symmetrize the
 problem, which physicists like very much, but also to introduce a well
 parametrized set of strategies, and more generally to provide a well defined
and workable family of models. 

In this game, $N$ agents have to choose one between two choices at each
time step; those who are in the minority win, the other
lose. Obviously, it is easier to loose than to win, as the number of winners cannot 
exceed that of the losers. If the game is played once,
only a random choice is reasonable, according to Game Theory~\cite{Game}. When the game is repeated, it is
sensible to suppose that agents will try to learn from the past in
order to outperform the other agents, hence, 
the question of learning arises, as the minority mechanism entails
a never-ending competition. 

Let me first introduce the game and the needed formalism. There are $N$
agents, agent $i$ taking action $a_i\in\,\{-1,+1\}$. A game master
aggregates the individual actions into $A=\sum_{i=1}^Na_i$ and gives
private payoffs $-a_ig(A)$ to each agent $i=1,\cdots,N$.  The minority structure of the
game implies that $g$ must be an odd function of
$A$. The simplest choice for $g$ may seem to be $g(A)=\sgn(A)$, but
 a linear function is better suited to mathematical analysis. The
MG is a negative sum game, as the total payoff given to the agents, is
$-\sum_{i=1}^Na_i g(A)=-g(A)A<0$, since $g$ is an odd
function. In particular, the linear payoff function gives a total loss
of $A^2$; when the game is repeated, the average total loss is nothing
else than the fluctuations of the attendance
$\sigma^2=\avg{A^2}$ where the average is over time.

From the point of view of the agents, it is a measure of payoff
wastage. That is why many papers on the MG consider it as the global
utility of the system (world utility hereafter), and try, of course to
minimize it (forward problem). I shall
review the quest for small $\sigma^2$, focusing on exact results,
 and show that all proposed mechanisms lead essentially to the same
results.\footnote{Evolutionary models~(see for instance
\cite{CZ97,CZ98,SavitEv1,SavitEv2}) are very different in nature, and
are not reviewed here, mostly because they are not exactly
solvable.}  A particular emphasis will be put on inductive behavior,
as it gives rise to particularly rich phenomenology while being well
understood. Finally, the reverse problem is addressed, by deriving what 
private payoff function $g$ to use given a world utility $W$ to minimize.

\section{No public information}

\subsection{``If it ain't broke, don't fix it''}

The arguably simplest behavior is the following \cite{Reents}:
if agent $i$ wins at time $t$, she  sticks to her choice
$a_i(t)$ until she looses, when she takes the opposite choice
with probability $p$. 
The dynamics is Markovian, thus can be solved exactly
\cite{Reents}. When $N$ is large,  the
fluctuations $\sigma^2$ are of order $(pN)^2$: indeed, as the number of losers is $\sim N$, 
the average number of agents changing their minds at time $t$ is $\sim pN$. 
Therefore, one can distinguish three regimes
\bit
\item $pN=x=cst$; this leads to small fluctuations
$\sigma^2=1+4x(1+x/3)$, which tend to the absolute minimum
$\sigma^2=1$ when
$x\to 0$. The time needed to reach the stationary state
is typically of order $\sqrt{N}$.
\item $p\sim 1/\sqrt{N}$; this yields $\sigma^2\sim N$, which is the
order of magnitude of produced by agents making independent choices.
\item $pN \gg 1$. In this case, a finite fraction of agents change
their mind at each time step and $\sigma^2=N(Np^2+4(1-p))/(2-p)^2\sim N^2$.
\eit

The major problem here is that $p$ needs to be tuned in order to reach
high efficiency. But it is very easy to design a feedback from the
fluctuations on $p$~\cite{Cunpub}, that lowers $p$ as long as the
fluctuations are too high, and to use the above results in order to
relate the fluctuations to $p(t\to\infty)$. Mathematically, this
amounts to take $p(t=0)=1$, $\dd p/\dd t =-f(p,N,t)$. For instance,
$f(t)=t^{-\beta}$ seems appropriate as long as $\beta$ is small
enough.  Note that $p(t)\to 0$ as $t\to\infty$, in words, the system
eventually freezes. From the optimization point of view, this is a
welcome, but as for agents, complete freezing, although being a Nash
equilibrium~\cite{Game}, is not satisfactory. as it may be better for
an agent sitting on the losing side to provoke an game-quake and to
profit from a re-arrangement of the winners/losers. Therefore, an
unanswered question is where to stop the time evolution of $p$.

Nevertheless, this simple example illustrates well what happens in MGs: the
efficiency essentially depends on the opinion switching rate, which
itself depends on the learning rate. It has to be small in order to
reach good level of efficiency.

\subsection{Inductive behavior}

Inductive behavior  can remedy the problems of the previous
learning scheme if, as we shall see, agents know the nature of the
game that they are playing. This subsection is a simplified version of
the simplest  setting for inductive agents of ref.~\cite{MC00}.
At time $t$, each agent $i=1,\cdots,N$ plays $+1$ with probability
$\pi_i(t)$, and $-1$ with probability $1-\pi_i(t)$. 
Learning consists in changing $\pi_i$ given the outcome of the game at
time $t$. For this purpose, each agent $i$ has a numerical  register $\Delta_i(t)$ 
which reflects her {\em perception} at time $t$ of the relative success of action
$+1$ versus action $-1$. In other words,  $\Delta_i(t)>0$ means that she believes
that action $+1$ has been more successful than $-1$. The idea is the
following: if agent $i$ observes $A(t)<0$ she will 
increase $\Delta_i$ and hence her probability of playing $a_i=+1$
at the next time step. Reinforcement here means that $\pi_i$ is an
increasing function of $\Delta_i$. For reasons that will become
obvious later, it is advisable to take $\pi_i=(1+m_i)/2$ and 
$m_i=\chi(\Delta_i)/2$, where $\chi$ is an
increasing function and $\chi(\pm\infty)=\pm1$.
The way in which $\Delta_i(t)$ is updated is the last and most crucial
element of the learning dynamics to be specified:
\be
\Delta_i(t+1)=\Delta_i(t)-\frac{1}{N}[A(t)-\eta a_i(t)].
\label{learn}
\ee
The $\eta$ term above describe the fact that
agent $i$ may account for her own contribution to $A(t)$. When
$\eta=0$, she believes that $A(t)$ is an {\em external process} on which
she has no influence, or does not know what kind of game she is
playing. She  may be called {\em naive} with this respect.
For $\eta=1$, agent $i$ considers only the behavior of
other agents $A_{-i}(t)=A(t)-a_i(t)$ and does not react to her own
action $a_i(t)$. As we shall see, this subtlety is the key
to high efficiency. The private utility of sophisticated agents
corresponds more or less to what is called Aristocrat Utility (AU) in COIN's nomenclature~\cite{COIN}.

\subsubsection{\bf Naive agents $\eta=0$}

It is possible to show that agents minimize the {\em predictability }
$H=\avg{A}^2$. As a consequence $H$ vanishes in the $t\to\infty$ limit.
There are of course many states with $H=0$ and the dynamics
selects that which is the ``closest'' to the initial
condition. To be more precise, let $\Delta_i(0)$ be the initial 
condition (which encodes the {\em a priori} beliefs of agent $i$ on which action
is the best one). As $t\to\infty$, $\avg{A}_t=\sum_i m_i(t)\to 0$ and 
$\Delta_i$ converges to
\be\label{dynUP1}
\Delta_i(\infty)=\Delta_i(0)+\delta A,~~~ \hbox{with}~~~
\delta A=\int_0^\infty d t\avg{A}_t.
\ee
The condition $\avg{A}_\infty=0$ provides an equation for $\delta A$
\be
0=\sum_{i=1}^N\chi\left(\Delta_i(0)+\delta A\right).
\label{eqnaive}
\ee
By the monotonicity property of $\chi$, this equation has one and
only one solution.

\begin{figure}
\centerline{\includegraphics[width=8cm,angle=270]{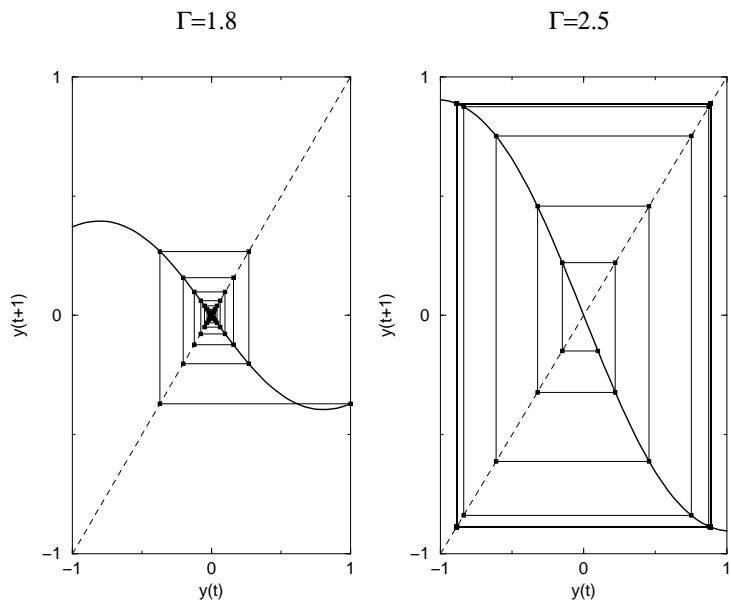}}
  \caption{Graphical iteration of the map $y(t)$ for $\Gamma=1.8<\Gamma_c$
and $\Gamma=2.5>\Gamma_c$}
  \label{figmap}
\end{figure}

The asymptotic state of this dynamics is information--efficient ($H=0$), 
but it is not optimal, as, in general, this state {\em is not} a Nash
equilibrium. The fluctuations are indeed determined by the behavior of
$\chi(x)$. This is best seen with a particular example: assume 
 that the agents behave according to a
Logit model of discrete choice~\cite{Logit} where the probability of choice $a$ is
proportional to the exponential of the ``score'' $U_a$ of that choice:
$\pi(a)\propto e^{\Gamma U_a/2}$. With only two choices $a=\pm 1$, 
$\pi(a)=(1+am)/2$ and $\Delta=U_+-U_-$, we obtain
\footnote{This learning model has been introduced
by \cite{Cavagna2} in the context of the MG.}
\be
\chi(\Delta)=\tanh (\Gamma\Delta), ~~~~\forall i.
\label{logit}
\ee
Here $\Gamma$ is the learning rate, which measures the scale of 
the reaction in agent's behavior (i.e. in $m_i$) to a change
in $\Delta_i$~\cite{comment}.
We also assume that agents have no prior beliefs: $\Delta_i(0)=0$.
Hence $\Delta_i(t)\equiv y(t)/\Gamma$ is the same for all agents.
From the results discussed above, we expect, in this case the 
system to converge to the symmetric Nash equilibrium $m_i=0$ 
for all $i$. This is not going to be true if agents are too reactive,
i.e. if $\Gamma>\Gamma_c$. Indeed, $y(t)=\Gamma\Delta_i(t)$ satisfies the equation
\bea
y(t+1)&=& y(t)-\frac{\Gamma}{N}\sum_{i=1}^N a_i(t)\nonumber\\
&\simeq & y(t)-\Gamma\tanh[y(t)]
\label{dyn0}
\eea
where the approximation in the last equation relies on the law
of large numbers for $N\gg 1$. Eq. \req{dyn0} is a dynamical system. 
The point $y^0=0$ is stationary,
but it is easy to see that it is only stable for $\Gamma<\Gamma_c=2$.
For $\Gamma>2$, a cycle of period $2$ arises, as shown in Fig.~\ref{figmap}. This has dramatic effects on the optimality of
the system. Indeed, let $\pm y^*$ be the two values taken by
$y(t)$ in this cycle\footnote{$\pm y^*$ are the two non-zero solutions of 
$2y=\Gamma\tanh(y)$.}. Since $y(t+1)=-y(t)=
\pm y^*$ we still have $\avg{A}=0$ and hence $H=0$.
On the other hand $\sigma^2=N^2 {y^*}^2$ is of order
$N^2$, which is even worse than the symmetric Nash equilibrium
$\pi_i=1/2$ for all $i$, where $\sigma^2=N$.

Hence, one finds again a transition from $\sigma^2\propto N$
to  $\sigma^2\propto N^2$ when the learning rate is too large.

\subsubsection{\bf Sophisticated agents $\eta>0$}

It is easy to check that with $\eta>0$, following the same
steps as in the previous section, the learning dynamics of agents
minimize the function
\be
H_\eta=\avg{A}^2-\eta\sum_{i=1}^N m_i^2,
\label{Heta}
\ee
Since $H_\eta$ is a harmonic function,
$H_\eta$ attains its minima on the boundary
of the domain $[-1,1]^N$. In other words, $m_i=\pm 1$ for all $i$ which
means that agents play pure strategies $a_i=m_i$. 
The stable states are optimal Nash equilibria for $N$ even. 
By playing pure
strategies agents minimize the second term of $H_\eta$. Of all corner
states where $m_i^2=1$ for all $i$, agents select those with 
$\avg{A}=0$ by dividing into two equal groups
playing opposite actions. All these states have minimal ``energy''
$H_\eta=-N\eta$. Which of these states is selected 
depends on the initial conditions $\Delta_i(0)$, but this  has no
influence on the outcome, since $\avg{A}=0$.

Note that
the set of stable states is {\em disconnected}. Each state has its
basin of attraction in the space of $\Delta_i(0)$. The stable
state changes discontinuously as $\Delta_i(0)$ is varied.
This contrasts with the case $\eta=0$ where Eq. \req{eqnaive} implies 
that the stationary state changes continuously with $\Delta_i(0)$
and the set of stationary states is connected.

For $N$ odd, similar conclusions can be found. 
This can be understood by adding a further agent to a state 
with $N-1$ (even) agents in a Nash equilibrium.
Then $H_\eta=(1-\eta)m_N^2$, so for $\eta<1$ the new agent
will play a mixed strategy $m_i=0$, whereas for $\eta>1$ it will 
play a pure strategy. In both cases other agents have no incentive
to change their position. In this case we find $\sigma^2\le 1$.


It is remarkable how the addition of the parameter $\eta$ 
radically changes the nature of the stationary state. 
Most strikingly, fluctuations are reduced by a factor $N$. From a
design point of view, this means that one has either to give a
personalized feedback to autonomous agents, or to make them more
sophisticated, for instance because they need to know the functional
form of the payoff.

\section{Public information}
As each agent has an influence on the outcome of the game, the
behavior of particular agent may introduce patterns that the other agents will try to 
exploit. For instance, if only one agent begins to think that 
the outcome of next game depends on some
external state, such as the present weather of Oxford, and behave accordingly,
then indeed, the outcome will depend on it.\footnote{This kind of
self-fulfilled prophecy is found for instance in financial markets,
where it is called `sunspot effect'.} But this means that other agents
can exploit this new pattern by behaving conditionally on the same
state.  One example of public information state family that agents may consider as
relevant is the past winning choices, for instance a window of size $M$ of past
outcomes~\cite{CZ97}. Each such state can be represented by a bit-string of size
$M$, hence there are $2^M$ possible states of the world. This kind of state has a
dynamics of its own: it diffuses on a so called De Bruijn Graph~\cite{CM01}. 
Another state dynamics consists simply in drawing at
random  the state at time $t$ from some
ensemble~\cite{Cavagna} of size $P$ (e.g. $P=2^M$). All exact results below are obtained with
this setup.

\subsection{Neural Networks}
Two types of neural networks have been studied in the context of the
MG \cite{MGNN1,MGNN2,Bak}. Beyond the mere academic question of how well or
badly they can perform, it is worth noting that these papers were interested for the first
time in {\em interacting} neural networks. 

Refs~\cite{MGNN1,MGNN2} introduced
simple perceptrons playing the minority game. Each perceptron $i=1,\cdots,N$ is made up of $M$ weights $\vec
w_i=(w_1^1,\cdots,w_1^M)$ which are drawn at random before the game
begins. The decision of network $i$ is $a_i=\sgn(\vec w.\vec \mu)$,
where $\vec \mu$ is the vector containing the $M$ last minority
signs. The payoff was chosen to be $-a_i\sgn(A)$. Neural networks are
trained following the usual Hebbian rule, that is, 
\be
\vec w_i(t+1)=\vec w_{i}(t)-\frac{\eta}{M}\vec\mu_t \sgn(A_t).
\ee
Under some simplifying assumptions, it is possible to find that the
fluctuations are given by \cite{MGNN1,MGNN2}
\be
\sigma^2=N+N(N-1)\pr{1-\frac{2}{\pi}\arccos\frac{K-1/(N-1)}{K+1}}
\ee
where $K=\frac{\eta^2\pi}{16}\pr{1+\sqrt{1+\frac{16(\pi-2)}{\eta^2\pi
N}}}$. The best efficiency, obtained in the limit $\eta\to 0$, is
given by
\be
\sigma^2=N\pr{1-\frac{2}{\pi}}.
\ee
This means that the fluctuations are at best of order $N$, and at
 worst of order $N^2$ when the learning rate is too high. This is
 likely to be corrected for neural networks with sophisticated
 private  utility.

\subsection{Inductive behavior}

El Farol's problem was introduced with public information and
inductive behavior~\cite{Arthur}, but with no precise characterization of the
strategy space. In most MG-inspired models, a strategy is a lookup table $a$,
or a map, or a function, which predicts the next
outcome $a^\mu$ for each state $\mu$, and whose entries are fixed for the
whole duration of the game. Each agent $i$ has a set of $S$ strategies, say
$S=2$ ($a_{i,1}$ and $a_{i,2}$), and use them essentially in the same way as before~\cite{CZ97}. 

\subsubsection{\bf Naive agents}

\begin{figure}
\centerline{\includegraphics[width=8cm]{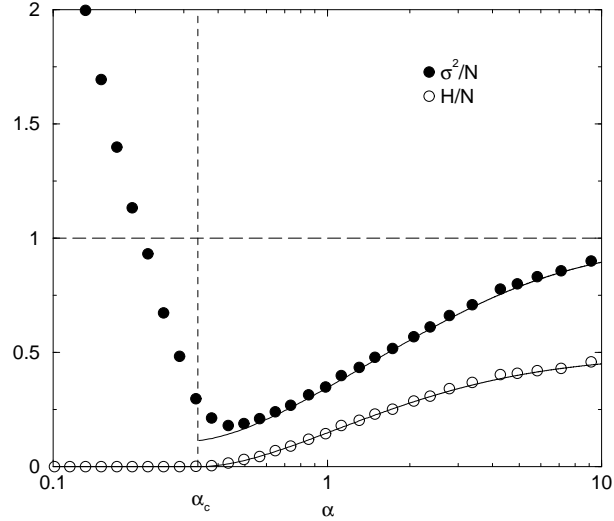}}
 \caption{Fluctuations and predictability produced by naive agents $P=64$,
  $300P$ iterations, average over $100$ realizations. This kind of
  figure is first found in ref.~\cite{Savit}.}
  \label{s2HRS}
\end{figure}

To each of her strategies, agent $i$ associate a score $U_{i,s}$ which
evolves according to \be\label{dynUP}
U_{i,s}(t+1)=U_{i,s}(t)-a_{i,s}^{\mu(t)}g[A(t)] \ee Since we consider
$S=2$, only the difference between $\Delta_i=U_{i,2}-U_{i,1}$ matters,
and \be
\Delta_{i}(t+1)=\Delta_{i}(t)-(a_{i,2}^{\mu(t)}-a_{i,1}^{\mu(t)})g[A(t)]
\ee Note that now $\Delta_i$ encodes the perception of the relative
performance of the two strategies of agent $i$, $\Delta_i>0$ meaning
that the agent $i$ thinks that strategy 2 is better than strategy 1,
and $m_i$ is the frequency of use of strategy $2$. As before, we
consider $\chi(x)=\tanh(\Gamma x)$. This kind of agents minimizes the
predictability, which has now to be averaged over the public
information states \be
H=\frac{1}{P}\sum_{\mu=1}^P\avg{A|\mu}^2=\ovl{\avg{A}^2} \ee where
$\ovl{Q}=\sum_{\mu=1}^PQ^\mu$ is a useful shortcut for the average
over the states of the world.  In contrast with the case with no
information, $H$ is not always canceled by the agents. This is due to
the fact that the agents are faced to $P$ possible states, but their
control over their behavior is limited: when they switch from one
strategy to another, they change their behavior potentially for all
states. In fact all macroscopic quantities such as $H/N$ and
$\sigma^2/N$ depend of the ratio
$\alpha=P/N$~\cite{Savit,CM99,CMZe99}, which is therefore the control
parameter of the system. Solving this model is much more complex and
requires tools of Statistical Physics of disordered
systems~\cite{dotsenko}. The resulting picture is that for infinite
system size ($P$, $N\to\infty$ with $P/N=\alpha={\rm
  cst}$)~\cite{CMZe99} (see also Fig~\ref{s2HRS}), \bit
\item  $H>0$ if $\alpha=N/P>\alpha_c=0.3374\ldots$. In this region, the system is not
informationally efficient. It tends to a stationary state which
is unique and stable, and does not depend either on $\Gamma$ or on
initial conditions. $\Gamma$ is a time scale~\cite{comment}.
\item $H=0$ when $\alpha<\alpha_c$. Since agents succeed in
minimizing $H$, the question for them is what should they do? They do
not known, and as a result, the dynamics of the system is
very complex: it depends on initial conditions\footnote{Physicists say
that it is not ergodic}~\cite{MC00,Oxf2,MC01,Coolen}, and on $\Gamma$~\cite{Cavagna2,MC01,Coolen}. Any value of the
fluctuations can be obtained, from $\sigma^2=1$ for very heterogeneous
initial conditions $\Delta_i(t=0)$ to $\sigma^2\sim N^2$ for $\Gamma=\infty$ and
homogeneous initial conditions, including $\sigma^2\sim N$ for
$\Gamma=0$ and any initial conditions. Two alternative theories
 have been proposed, one which is exact, but which has to be iterated 
\cite{Coolen}, and another one which rests
on a closed form for the fluctuations \cite{MC01}. Iterating the exact
theory is hard, since the $t-$th iteration is obtained by inverting
$t\times t$ matrices, and one has to average of several
realizations. Nevertheless, a hundred numerical iterations 
 bring promising results~\cite{Tobias}.
\eit
The origin of the phase transition can easily be understood in
terms of linear algebra: canceling $H=0$ means that $\avg{A|\mu}=0$ for
all $\mu$. This is nothing else than a set of $P$ linear equations of
$N$ variables $\{m_i\}$. As the
variables are bounded ($0\le m_i^2\le 1$), one needs more that $P$ of them,
$N=P/\alpha_c>P$ to be precise~\cite{MMM}.

In fact, the transition from low to high (anomalous) fluctuations does
not occur at $\alpha_c$ for finite system size as it clearly appears
on Fig~\ref{s2HRS}. This can be traced
back to a signal to noise ratio transition \cite{CM02}: 
the system is dynamically stable in the phase of $H>0$ as long as the signal to
noise ratio $H/\sigma^2$ is larger than $K/\sqrt{P}$ for some constant
$K$. This transition is universal for naive competing
agents. Hence in this kind of interacting agents systems, the ultimate cause of large
fluctuations is this signal-to-noise transition and high learning rate. Sophisticated agents
are not affected by this problem, as explained below.

\subsubsection{\bf Sophisticated agents}

\begin{figure}
\centerline{\includegraphics[width=8cm]{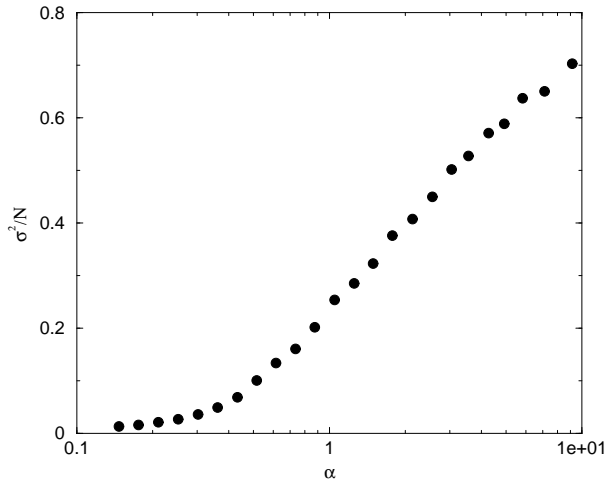}}
 \caption{Fluctuations produced by sophisticated agents $P=64$,
  $NIT=100P$, average over $100$ realizations.}
  \label{s2soph}
\end{figure}

As before, a sophisticated agent is able to disentangle her own contribution from $g(A)$. Eq \req{dynUP} becomes~\cite{CMZe99,MCZ}:
\be\label{dynUPsoph}
\Delta_{i}(t+1)=\Delta_{i}(t)-(a_{i,1}^{\mu(t)}-a_{i,2}^{\mu(t)})g(A(t)-a_i(t))
\ee
When the payoff is linear $g(A)=A$, the agents also minimize the fluctuations
$\sigma^2=\avg{A^2}$. Similarly, they end up using only one
strategy, which implies that $H=\sigma^2$. In this case, they cannot
cancel $A$ for all $\mu$ at the same time, hence
$\sigma^2/N>0$. How to solve exactly this
case is known in principle~\cite{CMZe99,MCZ}. 'In principle' here
means that the minimization of $\sigma^2$ is
hard from an technical point of view; how much harder is also
a question hard to answer. A first step was done in ref.~\cite{AM01},
which is able to describe reasonably well the behavior of the
system. Interestingly, in this case the signal-to-noise ratio
transition does not exist, as the signal is also the noise
($H=\sigma^2$), hence, there is no high volatility region (see Fig.~\ref{s2soph}). Therefore,
 the fluctuations are again considerably reduced by introducing
sophisticated agents. An important point here is that the number 
of stable final states $\{m_i\}$~\cite{AM01} grows exponentially when 
$N$ increases. Which one is selected depends on the initial conditions, but the
efficiency of the final state greatly fluctuates. As the agents (and
the programmer) have no clue of which one to select, the system ends
up having non-optimal fluctuations of order $N$, as seen of Fig.~\ref{s2soph}.

\section{Forward/reverse problems}

Inductive agents minimize a world utility whose determination is the first step
in solving the forward problem. Finding analytically its minimum is
then possible {\em in principle} thanks to methods of Statistical
Physics~\cite{dotsenko}. The reverse problem consists in starting from a
world utility $W$ and finding the appropriate private payoff. 

\subsection{Naive agents}

The case with no information ($P=1$) is trivial, since $\avg{A}=0$ in the
stationary state, hence all functions $H_{2n}=\avg{A}^{2n}$ ($n$
integer) are minimized
by a linear payoff.
When the agents have access to public information ($P>1$), the world utility $W$ given
any private payoff function $g(A)$ is~\cite{MC01}
\be\label{utility-naive}
W_{\rm naive}(\{m_{i}\})=\frac{1}{P}\sum_{\mu=1}^P\int_{-\infty}^{\,\infty}\,\frac{dx}{\sqrt{2\pi}}\e^{-x^2/2}\,G\pr{\avg{A|\mu}(\{m_{i}\})+x\,\sqrt{D}}
\ee
where $g(x)=\dd G(x)/\dd x$ and $D=\sigma^2-H=(N-\sum_i m_i^2)/2$. In other words, the agents select the set of
strategy usage frequencies $\{m_{i}\}$ that minimizes $U$. The final
state is unique and does not depend on the initial conditions. Note that
in Eq~\req{utility-naive}, only powers of $\avg{A|\mu}$
($\mu=1,\cdots,P$) appear, which
means that naive agents are only able to minimize world utility
 that only depend on these quantities. This
implies that a phase transition always happen if the agents are naive, and
even more, that it always happen at the same $\alpha_c=0.3374\ldots$,
as seen conjecture from numerical simulations in~\cite{SavitPayoff}. As explained above, $\alpha_c$ is the point
where it is algebraically possible to cancel all
$\avg{A|\mu}$~\cite{MMM}. The above theory also means that the stationary
state depends only weakly on the payoff, which can be seen numerically
by comparing the $m_i$ of a given set of agents for different
payoffs. 

The reverse problem is now to find $g$ given $W$. Let us focus on the
particular example $W=\ovl{\avg{A}^{2n}}$, ($n$ integer). 
First, one determines the world utility $W^{(2k)}$ associated with $g(x)=2k x^{2k-1}$, where $k$ is an integer,
\be
W^{(2k)}=\sum_{l=0}^{k}{2k \choose 2l}D^{k-l}X_{2(k-l)}H_{2l},
\ee
where $X_l=\int \exp(-x^2/2)x^l/\sqrt{2\pi}$ is the $l$-th moment of a
Gaussian distribution of unitary variance and zero average, and
$H_{2l}=\ovl{\avg{A}^{2l}}$ is the $2l-$norm of the vector 
$(\avg{A|\mu})$. Suppose now that one wishes to minimize
$W=H_{2n}$. This can be done {\em in principle} with a linear combination of
the $W^{(2k)}$
\be
W=\ovl{\avg{A|\mu}^{2n}}=\sum_{k=0}^{n}a_kW^{(2k)}=\sum_{k=0}^{n}a_k\sum_{l=0}^k{2k \choose 2l}D^{k-l}X_{2(k-l)}H_{2l},
\ee
The condition on the $\{a_k\}$ is that the coefficient of $H_{2k}$ be 0 for
$k=0,\cdots,n-1$, and the coefficient of $H_{2n}$ be 1, that is
\be
\sum_{m=k}^{n}a_m{2m\choose 2k} D^{m-k}X_{2(m-k)}=0~~~~1\le k\le n-1
\ee
and $a_n=1$. Then the problem is solved by finding the solution of these
$n-1$ linear equations of $a_k$, $k=1,\cdots,n-1$, and taking $g(x)=\sum_{k=1}^{n} a_k x^{2k-1}$. 
Note that
the set of the problems that naive agents can solve is of limited
practical interest.

\subsection{Sophisticated agents}

Sophisticated agents have instead
\be\label{utility-soph}
W_{\rm naive}(\{m_{i}\})=\frac{1}{PN}\sum_{\mu=1}^P\int_{-\infty}^{\,\infty}\,\frac{dx}{\sqrt{2\pi}}\e^{-x^2/2}\,\sum_iG\pr{\avg{A_{-i}^\mu(\{m_{i}\})}+x\,\sqrt{D_{-i}}}
\ee
where $D_{-i}=[(N-1)-\sum_{j\ne i}m_i^2]/2$. This case is much simpler than the previous one, as all agents
 end up playing only one strategy~\cite{CMZe99}, that is, $D_{-i}=0$. Therefore, in this case, if $g(A)=2kA^{2k+1}$, 
\be
W^{(2k)}=\avg{A^{2k}}.
\ee

 Interestingly, similar functions are well-studied in
Statistical Physics, where they usually represent the energy of
interacting magnetic moments called ''spins''~\cite{MPV}; a
(classical) spin can have two values $-1$ or $+1$, which is the
equivalent of choosing strategy $1$ or $2$.  A well-known qualitative
change occurs between $k=2$ and $k>2$, where the mathematical
minimization of $W$ is somehow less problematic; this may also be the
case in such MGs. The final state is not
unique, and depends on initial conditions, implying that agents cannot
are not particularly good at minimizing such functions.

\subsection{Example: agent-based optimization}

Some optimization problems are so hard to solve that they have a name:
they are hard, NP-hard~\cite{Garrey}.  There is no algorithm that can
find the optimum of this kind of problems in polynomial time. One of
them consists in finding amongst $N$ either analogic or binary
components the combination that is the least defective~\cite{CJ02}: in
the problem with analogic components, one has a set of $N$ measuring
devices; instead of $A$, each of them records the wrong value $A+a_i$
with a constant bias $a_i$, $i=1,\cdots,N$, drawn from a given
probability function. The problem is to find a subset such that the
average bias \be
\eps\{n_i\}=\frac{|\sum_{i=1}^Nn_ia_i|}{\sum_{j=1}^Nn_j} \ee is
minimal. Here $n_i=0,1$ depending on whether component $i$ is included
in the subset. Statistical Physics shows that $\avg{\eps_{\rm
    opt}}\sim C\, 2^{-N}/\sqrt{N}$ for large $N$, with $C\simeq 4.6$
(the average is over the samples). In order to find the optimal
subset, one cannot do better than enumerating all the $2^N$
possibilities. This makes it hard to tackle such problems for $N$
larger than $40$ with nowadays computers. Agent-based optimization on
the other hand needs typically $O(N)$ iterations and can be used with
much larger samples. It is clear that one cannot expect this method to
perform as well as the enumeration, still how well it perform as a
function of the setup is a valuable question. Ref~\cite{Kagan}
compares a set of private payoffs and concludes that agent-based
optimization is better than simulated annealing for short times and
large samples, provided that the agents' private utility is
``aristocratic''.

\begin{figure}
\centerline{\includegraphics[width=8cm]{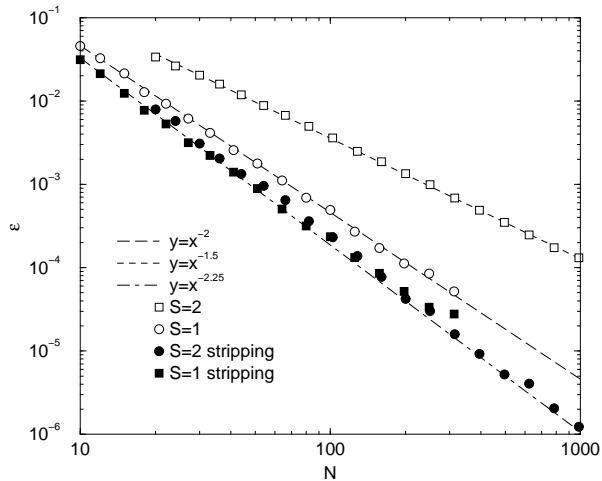}}
  \label{S1}
  \caption{Average error $\eps$ versus the size $N$ of the defective
  component set for MG with $S=2$ (circles), and $S=1$ (squares),
  $S=2$ with removal (stars) and $S=1$ with removal (full
  squares. $500N$ iterations per run, averages over 1000 samples.}

\end{figure}

Optimizing $h=|\sum_{i=1}^Nn_ia_i|$ and then dividing by the number of
components used in the chosen subset leads to almost optimal
subsets~\cite{CJ02}. Hence, we can use sophisticated MG-agents in
order to optimize $h^2$~\cite{CJ02unpubl}, which plays the role of the
fluctuations in the MG. The most straightforward application of the MG
is to give two devices to each agents, which are their strategies.
Each agent ends up playing with only one strategy. This setup
constraints the use of $N/2$ devices in the optimal subset, and gives
an error of order $N^{-1.5}$, to be compared with the exponential
decay of the optimal average error $\eps_{\rm opt}$. One can
unconstrain the agents by giving only one component to each agent, and
letting them decide whether to include their components or not into
$\eps$, making the game 'grand canonical'~\cite{SZ99,J99}. This is
achieved by the following score evolution \be
U_i(t+1)=U_i(t)-a_i[A-n_i(t)a_i] \ee and $n_i(t)=\Theta[U_i(t)]$. The
$-n_ia_i$ term makes the agents sophisticated.  This gives similar
results as those of ref~\cite{Kagan}, as
indeed the Aristocrat Utility is essentially the same concept as
sophisticated agents. But in any case, it minimizes the fluctuations,
but does not optimize them. The resulting error $\eps$ is much better
with $S=1$ than with $S=2$: it decays $\sim N^{-2}$ (Fig~\ref{S1}).
Therefore, as in the optimal case, unconstraining the problem by not
fixing the number of selected components leads to much better
efficiency.

At this stage, one can improve substantially the error, still
remaining in the $O(N)$ complexity regime.  First, since the agents
update their behavior simultaneously, they may be unable to
distinguish whether removing only one component improves the error. We
can do it by hand at the end of the simulations, repeatedly. This is a kind of greedy algorithm. On
average, about 1.5 components are removed.  In both the $S=2$ and
$S=1$ cases, this results into a large improvement (see
Fig.~\ref{S1}), and curioulsy produces the same error, with a decay
$\sim N^{-2.3}$. Nevertheless, the final error is still far from
optimality. This illustrates how hard this optimization problem is.
Much better results can be obtained by removing a group of 2, or 3
components, {\em ad libitum}, but of course, this needs much more
computing resources ($O(N^2)$, $O(N^3)$, \dots), and eventually
amounts to enumerating all possibilities.

\begin{figure}
\centerline{\includegraphics[width=8cm]{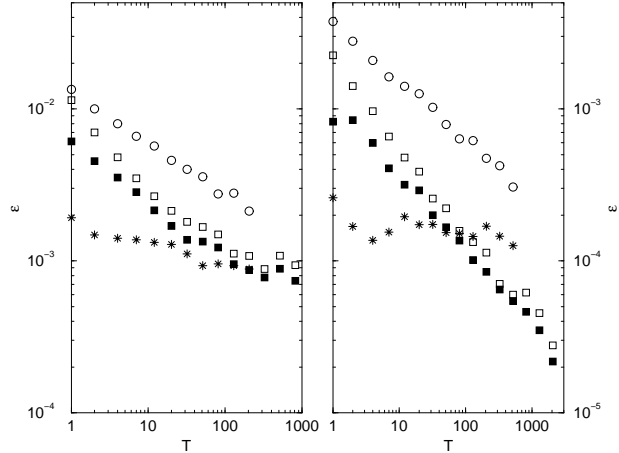}}
  \caption{Average error $\eps$ versus the number of runs for each sample of the defective
  component set for MG with $S=2$ (circles), and $S=1$ (squares),
  $S=2$ with removal (stars) and $S=1$ with removal (full squares). Left panel: $N=20$, averages over 1000
  samples; right panel: $N=50$, average over 200 samples.}
  \label{Ntry}
\end{figure}

Here is the second trick that keeps the complexity with the $O(N)$
regime.  As mentioned, the final state depends on the initial
conditions, and is often not optimal or not even near optimal. But it
is still a local minimum of $h^2$. Therefore the idea is to do $T$
runs with the same set of defective devices, changing the initial
condition $U_i(t=0)$, and to select the best run. It is a kind of
simulated annealing~\cite{Kirk} with zero temperature, or partial
enumeration where repetition would be allowed. Interestingly,
Figure~\ref{Ntry} reports that the decay is apparently a power-law
first, and then begins to saturate. For $S=1$, the exponent is about
$-0.5$, and $0.4$ for $S=2$; it depends weakly on $N$. Remarkably, the
error decreases faster with $S=1$ agents than $S=2$. Note that the
optimal value is at about $10^{-6}$, hence, agents are far from it.
This is due to the fact that the agents use too many components.
Nevertheless, the improvement brought by this methods is impressive,
and increases as $N$ increases, but cannot keep up with the
exponential decay of $\eps_{\rm opt}$: the difference becomes more and
more abysmal. The component removal further lowers the error (same
figure), and more in the $S=2$ that in the $S=1$ one. This advantage
is reversed for $T$ large enough when $N$ is larger, as reported by
the right panel Fig.~\ref{Ntry}).
 
The other optimization problem recycles binary components~\cite{CJ02}:
one has a set of $N$ partially defective processors, each of them able
to perform $P$ different operations. The manufacturing process is
supposed to be fault with probability $f$ for each operation of each
component.  Mathematically, the operation $\mu$ of processor $a$ is
permanently defective ($a^\mu=-1$) with probability $f$ and works
permanently with probability $1-f$ ($a^\mu=1$). The probability that a
component is working becomes vanishingly small when $P$ grows at fixed
$f$.  The task consists in finding a subset such that the majority of
its components gives the right answer, that is, 
\be 
\sum_{i=1}^N n_i a_i^\mu>0~~~~{\rm for~all}~~\mu=1,\cdots,P 
\ee
\begin{figure}
\centerline{\includegraphics[width=8cm]{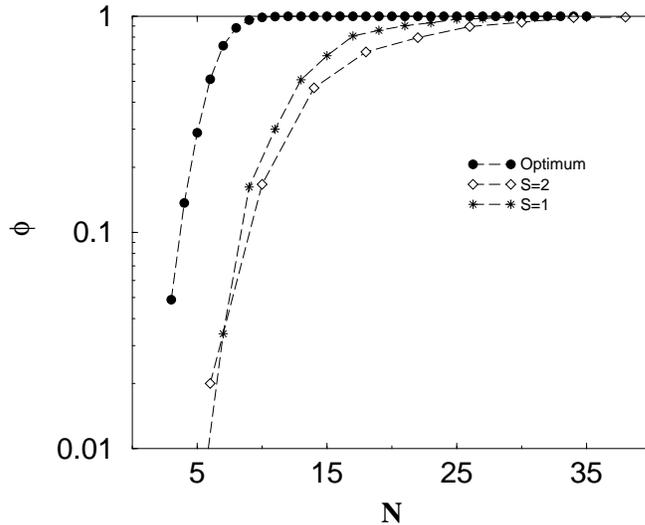}}
  \caption{Fraction of samples for which a perfectly working
  subset of components can be found.  $f=0.2$, average over 1000 runs.}
  \label{figphi}
\end{figure}
Surprisingly, the fraction $\phi$ of samples in which a perfectly working
subset of components can be found increases very quickly as $N$ grows
at fixed $P$ and $f$~\cite{CJ02} (see also fig.~\ref{figphi}).
Finding a subset that perfectly works is an easy problem when it is possible,
but finding the one which has the least components is a hard
problem~\cite{CJ02}. By contrast with the minimization of fluctuations,
here one wishes to maximize $A$ given $\mu$, that is, the
predictability $H$. Since all the agents eventually use only one strategy in majority
games~\cite{CM99}, $H=\sigma^2$, hence, the fluctuations $\sigma^2$
are also maximized: naive agents are also sophisticated in this case. A simple majority game
does not favor any particular sign of $A^\mu$ {\em a priori}. However, if $f\ll1/2$ the sign
$+$, hence mostly working combinations, are favored. In practice, a majority game payoff increase is
$ag(A)$ instead of $-ag(A)$ as in minority games, which means that
here one has
\be
U_i(t+1)=U_i(t)+a_i^{\mu(t)}A(t)
\ee
Majority games with $S=1$ turn out to be better than  those
agents with $S=2$, as shown in Fig.~\ref{figphi}, where the results
of enumeration are also displayed. As the problem to find a
working subset is easy for $N$ large enough, the agents are successful.

\section{Conclusions}

The efficiency of Minority Games seems to be universal with respect to
agents' learning rate: if the latter is too high, anomalous
fluctuations, hence small efficiency arise. However, these are totally
suppressed if the agents are sophisticated, who can optimally
coordinate if there is no public information.  An unexplored issue is
what happens with neural networks taking into account their impact on
the game. Based on this `universality', it would be tempting to study
neural networks with the sophisticated payoff.

The study of forward/reverse problems showed the limitations of
agent-based optimization in hard cases, which leaves the interesting
open question of how to improve the overall performance, and how the
setup of agent-based models can and must be tuned for individual
cases.

I am grateful to J.-P. Garrahan, N. F. Johnson, M. Marsili, D.
Sherrington and Yi-Cheng Zhang for numerous discussions. This work has
been supported by EPSRC.

\end{document}